\documentstyle[prd,aps,preprint,tighten,eqsecnum]{revtex} 
\begin{document} 
\preprint{hep-th/9702158} %places preprint number at 
%%                         top right of title page 
%\draft %omission stifles PACS numbers 
\title{Duality Invariance of Black Hole Creation Rates}  
\author{J. David Brown} 
\address{Department of Physics and Department of Mathematics,\\ 
   North Carolina State University, Raleigh, NC 27695--8202}
%\date{\today} 
\maketitle 
%%%%%%%%%%%%%%%%%%%%%%%%%%%%%%%%%%%%%%%%%%%%%%%%%%%%%%%%%%%%%%%%%%%%%%%%%%%%%%%%
\begin{abstract} 
Pair creation of electrically charged black holes and its dual process, pair 
creation of magnetically charged black holes, are considered. It is shown 
that the creation rates are equal provided the boundary conditions for the 
two processes are dual to one another. This conclusion follows from a careful 
analysis of boundary terms and boundary conditions for the Maxwell action. 
\end{abstract} 
\pacs{???} 
%%%%%%%%%%%%%%%%%%%%%%%%%%%%%%%%%%%%%%%%%%%%%%%%%%%%%%%%%%%%%%%%%%%%%%%%%
%%%%%%%%%%%%%%%%%%%%    CHAPTER ONE    %%%%%%%%%%%%%%%%%%%%%%%%%%%%%%%%%%
%%%%%%%%%%%%%%%%%%%%%%%%%%%%%%%%%%%%%%%%%%%%%%%%%%%%%%%%%%%%%%%%%%%%%%%%%
\section{Introduction} 
The classical Einstein--Maxwell equations  are invariant under a duality 
transformation in which the electric and magnetic fields exchange  
roles. Recently the subject of electromagnetic duality has arisen in 
the context of black hole pair creation.  One can consider 
the pair creation of magnetically charged black holes 
and its ``dual process", pair creation of electrically charged black 
holes.\footnote{Gibbons \cite{Gibbons} first 
recognized that the pair creation 
rate for magnetically charged black holes can be computed in the semiclassical
approximation (or, more precisely, the zero--loop approximation) using 
instanton
techniques. The analysis was presented in Ref.~\cite{GS}. The  technical 
details of the calculation for
electrically charged  black holes were laid out in Ref.~\cite{PairCreation},
although an absolute creation rate was not computed there---only the rate
relative to  those for charged matter distributions.}
Are the creation rates equal? The answer is yes,\footnote{An explicit
demonstration was first given by Hawking and  Ross \cite{HR} and more recently 
by Deser, Henneaux, and Teitelboim \cite{DHT}.} as one might expect on 
physical grounds, but it is not immediately obvious from a technical 
standpoint. In fact, the factor $F_{\mu\nu}F^{\mu\nu} = 2(B^2 - E^2)$ in 
the Maxwell Lagrangian changes  sign if the electric and magnetic fields 
are exchanged. Therefore it might appear that the  semiclassical expressions 
for the two creation rates, which are both   proportional to the exponential 
of the action, should include an exponential suppression in one case and 
an exponential enhancement in the other. However, as Hawking and Ross 
\cite{HR} recognized, boundary conditions play a crucial role in the 
analysis. One can see the importance of 
boundary conditions by recalling that the choice of boundary conditions 
affects the boundary terms that appear in the action. Hawking 
and Ross \cite{HR} did not use this observation in their primary argument 
for the equality of the pair creation rates, but they did give a brief, 
secondary argument based on the analysis of boundary terms in the Maxwell 
action. The main purpose of this paper is to elaborate on this argument by 
presenting a  complete and detailed analysis of boundary terms and boundary 
conditions for the electromagnetic field action.

Section 2 begins with an analysis of the action, denoted
$S_1[g,F]$, which is the integral of the Maxwell Lagrangian. I show that the
boundary conditions for $S_1$ include fixation of the component $B^\vdash$ 
of the magnetic field normal to the system boundary, and the components $E^a$ 
of the electric field tangent to the system boundary. Let us assume that for 
a certain choice of boundary values $B^\vdash$ and
$E^a$, there exists an instanton  (such as the magnetic Ernst solution) 
describing
pair creation of magnetically charged black holes. The action $S_1$ can be 
used to compute the creation rate under the given set of boundary conditions. 
Now consider
the pair creation of electrically charged black holes.  An instanton (such as 
the
electric Ernst solution) describing  the pair creation of electrically 
charged black
holes certainly exists for {\em some} choice of boundary values $B^\vdash$ 
and $E^a$,
and $S_1$ can be used to find the creation rate for electrically charged 
black holes
under these conditions. However, we wish to compare the creation rates 
between the
magnetic and electric cases, and for this comparison to be meaningful the 
boundary
conditions must be related appropriately. In particular, the creation rate 
for
magnetically charged black holes under the conditions of fixed $B^\vdash$ 
and $E^a$
should be compared to the creation rate for electrically charged black holes 
under the
``dual conditions" of fixed
$E^\vdash$ and $B^a$. The action $S_1$ is not appropriate for these dual 
boundary
conditions. Thus, we are led to consider another electromagnetic field 
action,
$S_2[g,F]$, which differs from $S_1$ by a boundary term. The boundary 
conditions for
$S_2$ include fixation of the normal component $E^\vdash$ of the
electric field and the tangential components $B^a$ of the magnetic field. 

The key result, derived in Sec.~2, is that the electromagnetic field actions
$S_1$ and $S_2$ satisfy the relationship $S_{1}[\tilde g,\tilde F] 
= S_{2}[\tilde
g,-{}^*{\tilde F}]$. Here, 
$g_{\mu\nu} = {\tilde g}_{\mu\nu}$, $F_{\mu\nu} = {\tilde F}_{\mu\nu}$ 
is a given classical solution and $g_{\mu\nu} = {\tilde g}_{\mu\nu}$,
$F_{\mu\nu} = -{}^{*}{\tilde F}_{\mu\nu}$ is the dual solution. 
This observation solves the pair creation problem: 
The pair creation rate for magnetically 
charged black holes under the conditions of fixed $B^\vdash$ and $E^a$
is proportional to the exponential of
$S_1$ evaluated  at an appropriate classical solution $\tilde g$, $\tilde F$ 
(such as 
the magnetic Ernst instanton). The pair creation rate for electrically 
charged black 
holes under the dual conditions of fixed $E^\vdash$ and $B^a$ is proportional 
to the
exponential of $S_2$ evaluated at the dual solution 
$\tilde g$, $-{}^*\tilde F$ (the
electric Ernst instanton). From the relationship $S_{1}[\tilde g,\tilde F] =
S_{2}[\tilde g,-{}^*{\tilde F}]$, it follows that the pair creation rates 
are equal in the zero--loop approximation.\footnote{The argument presented 
in Ref.~\cite{DHT} for the equality of the pair creation rates is rather different.
There, the authors use Hamiltonian  methods to construct an action in which both the
electric charge and the  magnetic charge are fixed at spatial infinity, and only the
transverse parts  of the electric and magnetic fields are dynamical. They then 
show that this  action is invariant under a duality transformation in which,
simultaneously,  the transverse electric and magnetic fields and the electric and
magnetic  charges are transformed.} 
 
Section 2 concludes with an argument that shows the impossibility of fixing 
$E^\vdash$
and $E^a$, or $B^\vdash$ and $B^a$, as boundary data. Thus, one cannot speak 
of black
hole pair creation (or any other physical process) under such conditions. 

%%%%%%%%%%%%%%%%%%%%%%%%%%%%%%%%%%%%%%%%%%%% 
\section{Electromagnetic Field Actions and Black Hole Pair Creation} 
Consider the spacetime manifold ${\cal M}$ 
with metric $g_{\mu\nu}$.  Assume for the moment that ${\cal M}$ 
is the cross product of a space manifold $\Sigma$ 
and the time line interval $I$, ${\cal M}=\Sigma\times I$. The boundary of
$\Sigma$ is $\partial\Sigma$, which need not be simply connected.
The boundary of ${\cal M}$ consists of the 
surfaces $\Sigma'$ and $\Sigma''$ at the endpoints of the line interval 
and the history 
${\cal T}=\partial\Sigma\times I$ of the spatial boundary.\footnote{The 
spatial boundary can be chosen at asymptotic infinity for those cases in 
which asymptotic infinity exists. Here the spatial boundary is left 
unrestricted, so the analysis applies whether or not an asymptotic 
region exists.} 

Let $\gamma^\mu_i$ denote the projection mapping  
$T^*_p{\cal M} \to T^*_p{\cal T}$, where $p$ is a point in ${\cal T}$. 
(In terms of local coordinates $X^\mu$ on ${\cal M}$ and $x^i$ on ${\cal T}$,  
$\gamma^\mu_i = \partial X^\mu(x,r)/\partial x^i$ where ${\cal T}$ is an 
$r={\rm const}$ surface.) Also define $\gamma_{\mu}^{i} =
g_{\mu\nu}\gamma^{ij}\gamma^{\nu}_{j}$ where $\gamma_{ij} 
= \gamma^\mu_i g_{\mu\nu} \gamma^\nu_j$ is the induced metric on ${\cal T}$ 
and $\gamma^{ij}$ is its inverse. With these definitions the 
identity tensor at points $p\in{\cal T}$ becomes $\delta^{\nu}_{\mu}= 
n_{\mu}n^{\nu} + \gamma_{\mu}^{i} \gamma^{\nu}_{i}$ where 
$n_{\mu}$ is the spacelike unit normal of ${\cal T}$. 
Likewise, let $h^\mu_{\underline i}$ denote the projection mapping  
$T^*_q{\cal M} \to T^*_q{\Sigma''}$, where $q$ is a point in $\Sigma''$. 
Define $h_{\mu}^{\underline i} = 
g_{\mu\nu}h^{\underline{ij}}h^{\nu}_{\underline j}$ where $h_{\underline{ij}} 
= h^\mu_{\underline i} g_{\mu\nu} h^\nu_{\underline j}$ is the induced metric 
on $\Sigma''$ and $h^{\underline{ij}}$ is its inverse. The identity tensor 
at points $q\in\Sigma''$ is $\delta_{\mu}^{\nu}= 
-{\bar u}_{\mu} {\bar u}^{\nu} + h_{\mu}^{\underline i} 
h^{\nu}_{\underline i}$ where 
${\bar u}_{\mu}$ is the timelike unit normal of $\Sigma''$. Similar 
definitions hold for $\Sigma'$. 

Now consider the action for the electromagnetic field,\footnote{For 
simplicity, I define the action as the argument of the exponential in the 
path integral. Hence the factor of $i$ in Eq.~(1).}
\begin{equation}
	S_1[g,F] = -\frac{i}{16\pi}\int_{\cal M} d^4X \sqrt{-g} 
    F^{\mu\nu}F_{\mu\nu} \ ,\eqnum{1}
\end{equation} 
where $F_{\mu\nu} = 2\partial_{{\scriptscriptstyle [}\mu} 
A_{\nu{\scriptscriptstyle ]}}$. 
The notation $S_1[g,F]$ indicates only that $S_1$ is a functional of the 
metric tensor $g_{\mu\nu}$ and the electromagnetic field strength 
$F_{\mu\nu}$; it is not meant to imply that 
$F_{\mu\nu}$ is a fundamental variable in the variational
principle.  Rather, $S_1$ is to be varied with respect to
the potential $A_\mu$ and the metric $g_{\mu\nu}$. The variation of $S_1$ 
with respect to $A_\mu$ yields the source free Maxwell equations 
$\nabla_\mu F^{\mu\nu} = 0$ and the variation of $S_1$ with respect to
$g_{\mu\nu}$ yields the stress--energy--momentum tensor for the 
electromagnetic field. 

The boundary conditions suitable for the variational principle
based on the action $S_1$ are determined from the variation 
\begin{equation} 
   \delta S_1 = {\hbox{(eom's)}} 
   + \frac{i}{4\pi}\int_{\Sigma'}^{\Sigma''} d^3x 
   \sqrt{h} {\bar u}_\mu F^{\mu\nu} 
   \delta A_\nu   - \frac{i}{4\pi} \int_{\cal T} d^3x \sqrt{-\gamma} 
   n_\mu F^{\mu\nu} 
   \delta A_\nu \ .\eqnum{2}
\end{equation} 
Here, (eom's) are terms that yield the classical equations of motion. 
Inserting the identity tensor between the factors $F^{\mu\nu}$  and 
$\delta A_\nu$, we find 
\begin{equation} 
   \delta S_1 = {\hbox{(eom's)}} 
   + \frac{i}{4\pi}\int_{\Sigma'}^{\Sigma''} d^3x \sqrt{h} {\bar u}_\mu 
   F^{\mu\nu} 
   h_{\nu}^{\underline i} \delta A_{\underline i} 
   - \frac{i}{4\pi}\int_{\cal T} d^3x \sqrt{-\gamma} n_\mu F^{\mu\nu} 
   \gamma_{\nu}^{i} \delta A_i \eqnum{3}
\end{equation} 
where $A_{\underline i} = A_{\nu} h^{\nu}_{\underline i}$ and $A_{i} = 
A_{\nu} \gamma^{\nu}_{i}$. (This derivation uses the result 
$\delta\gamma^\mu_i =
0$ for variations in the dynamical variables $g_{\mu\nu}$ and $A_\mu$. Also 
note that
$\delta n_\mu$ and $\delta h_\mu^{\underline i}$ are each proportional to 
$n_\mu$. 
Similar results hold for $h^\mu_{\underline i}$, $h_\mu^{\underline i}$, 
and ${\bar u}_\mu$.) Equation (3) shows that, for the action 
$S_{1}$, the boundary conditions consist of fixed $A_{\underline i}$ 
on $\Sigma'$ and $\Sigma''$, and fixed $A_{i}$ on ${\cal T}$. 

Let us see what these boundary conditions imply for the electric and magnetic 
fields. First consider $\Sigma'$  and $\Sigma''$. The electric and magnetic 
fields measured by observers with  four--velocities ${\bar u}^\mu$ at 
$\Sigma'$ or $\Sigma''$ are 
$E^\mu = F^{\mu\nu}{\bar u}_\nu$ and $B^\mu = -{}^\star F^{\mu\nu} 
{\bar u}_\nu$, respectively. Here, ${}^\star F^{\mu\nu} = 
(1/2)\epsilon^{\mu\nu\rho\sigma}F_{\rho\sigma}$ is the dual of the field 
strength $F^{\mu\nu}$, where $\epsilon_{\mu\nu\rho\sigma}$ is the 
spacetime volume form. Expressed as a vector in the tangent space to 
$\Sigma'$ or $\Sigma''$,  the magnetic field $B^{\underline i} = 
h^{\underline i}_\mu B^\mu$ becomes 
\begin{equation} 
   B^{\underline i} = \epsilon^{\underline{ijk}} \partial_{\underline j} 
   A_{\underline k} \eqnum{4} 
\end{equation} 
where $\epsilon_{\underline{ijk}}$ is the 
volume form on $\Sigma'$ or $\Sigma''$. (This derivation uses the result 
$h_{\underline i}^\mu h_{\underline k}^\nu \partial_{{\scriptscriptstyle 
[}\mu}
h^{\underline k}_{\nu{\scriptscriptstyle ]}} = 0$ and the fact that each 
term in
$\partial_{{\scriptscriptstyle [}\mu} {\bar u}_{\nu{\scriptscriptstyle ]}}$ 
is proportional to either 
${\bar u}_\mu$ or ${\bar u}_\nu$.) Recall that for $S_1$,
$A_{\underline i}$ is fixed on the boundary elements $\Sigma'$ and 
$\Sigma''$. The result (4) shows that the magnetic field 
$B^{\underline i}$ is fixed on $\Sigma'$ and 
$\Sigma''$. (It is assumed that the induced metric $h_{\underline{ij}}$ 
and hence also the volume form $\epsilon_{\underline{ijk}}$ are fixed on 
$\Sigma'$ and $\Sigma''$.) 

Now turn to the boundary element ${\cal T}$ with spacelike unit normal 
$n^\mu$. 
For the considerations that follow the spacetime manifold $\cal M$ need not 
be a
product $\Sigma\times I$. The boundary element $\cal T$ is the interface 
between the
system and the exterior universe. It is here that boundary conditions are 
imposed for
such problems as black hole pair creation. Conceptually, the boundary  
values at
${\cal T}$ are considered to be fixed and/or monitored as the system evolves 
in time 
by a fleet of observers located at the system boundary. These observers must 
synchronize their measurements in time, which implies the existence of a 
foliation of
$\cal T$ into $t={\rm const}$ hypersurfaces $\cal B$. (If $\cal M$ is a
product $\Sigma\times I$, we can identify ${\cal B}$ with $\partial\Sigma$). 
I will assume that the observers are at rest in these time 
slices.\footnote{This is not a
necessary assumption, just a convenient one.  Alternatively one can consider 
observers
who are boosted relative to the constant time surfaces \cite{BLY}.}  Let us 
denote the
four--velocities of the  $\cal T$ observers by $u^i$; as vectors in 
$T_p{\cal M}$
with $p\in\cal T$, the four--velocities are $u^\mu = u^i\gamma_i^\mu$. 
Let $\sigma^i_a$
denote  the projection mapping $T^*_{p}{\cal T}\to T^*_p{\cal B}$. (In terms 
of local 
coordinates $x^i$ on $\cal T$ and $z^a$ on $\cal B$, $\sigma^i_a = \partial
x^i(z,t)/\partial z^a$ where $\cal B$ is a $t={\rm const}$ surface.) Also 
define 
$\sigma^a_i = \gamma_{ij}\sigma^{ab}\sigma_b^j$ where $\sigma_{ab} = 
\sigma^i_a
\gamma_{ij} \sigma^j_b$ is the induced metric on $\cal B$ and $\sigma^{ab}$ 
is its inverse.  With these definitions  the identity tensor on
$\cal T$ becomes $\delta_i^j = -u_i u^j + \sigma_i^a\sigma_a^j$. 

The electric and magnetic 
fields measured by the observers $u^i$ are 
$E^\mu = F^{\mu\nu}{u}_\nu$ and $B^\mu = -{}^\star F^{\mu\nu} {u}_\nu$, 
respectively. A short calculation shows that the component of 
$B^{\mu}$ orthogonal to ${\cal T}$, namely $B^\vdash = B^{\mu}n_{\mu}$, is 
\begin{equation}
	B^\vdash = \epsilon^{ab}  \partial_{a}A_{b}
	\ ,\eqnum{5}
\end{equation}
where $\epsilon_{ab}$ is the volume form on ${\cal B}$ and $A_a = 
A_i\sigma^i_a = 
A_\mu\gamma^\mu_i\sigma^i_a$ are the components of the electromagnetic 
potential
tangent to $\cal B$. Recall that for the action $S_1$, $A_i$ (and hence 
also $A_a$)
is fixed on the boundary element $\cal T$. The result (5) shows that the 
fixed boundary data  includes the component $B^\vdash$ of the magnetic 
field that is
orthogonal  to the system boundary $\cal B$ as measured by the observers 
$u^i$. (It is assumed  that the induced metric $\gamma_{ij}$ is fixed on 
${\cal T}$.) Another short calculation shows that the components of the
electric field  tangent to ${\cal B}$, namely $E^{a} = 
E^{\mu} \gamma_{\mu}^{i} \sigma_{i}^{a}$, are given by 
\begin{equation}
	E^{a} = 2 \sigma^{a}_{i} \gamma^{ij} u^{k} 
	\partial_{{\scriptscriptstyle [}j} A_{k{\scriptscriptstyle ]}} 
	\ .\eqnum{6} 
\end{equation}
Thus the fixed boundary data for $S_1$ also includes the components $E^a$ of 
the electric field tangent to the system boundary ${\cal B}$ as measured by 
the observers $u^i$. 

The magnetic charge $Q_m$ for the system is defined by the magnetic field 
flux through a surface $\cal B$: 
\begin{equation} 
  4\pi Q_m = \int_{{\cal B}} {\mathbf{F}} 
  = \int_{{\cal B}} d^2z\sqrt{\sigma}
   B^\vdash 
  \ .\eqnum{7}
\end{equation} 
The result (5) shows that $Q_m$ is fixed as boundary data in the variational 
principle for $S_1$. Of course, there are no local 
magnetic field sources in the system since the Maxwell equations imply the 
vanishing of the (covariant in space $\Sigma$) divergence of $B^\mu$.  
However, the charge $Q_m$ need not be zero because $\cal M$ can have a 
nontrivial topology. In particular, black holes can carry magnetic charge. 

The action $S_2$ for the electromagnetic field is defined by 
\begin{eqnarray}
	S_2[g,F] & = & S_1[g,F] + \frac{i}{4\pi}\int_{\cal M} d^4X\, 
	\partial_\mu(\sqrt{-g} 
	F^{\mu\nu} A_\nu) \nonumber\\ 
	& = & \frac{i}{16\pi}\int_{\cal M} d^4X \Bigl\{ \sqrt{-g} 
    F^{\mu\nu}F_{\mu\nu}  + 4 \partial_\mu(\sqrt{-g} F^{\mu\nu}) A_\nu 
    \Bigr\} \ .\eqnum{8} 
\end{eqnarray} 
Note that $S_2$ differs from $S_1$ by a boundary term. 
Like $S_1$, the fundamental variables in $S_2$ are
$A_\mu$ and $g_{\mu\nu}$. The variation of $S_2$ is given by 
\begin{eqnarray} 
   \delta S_2 & = & {\hbox{(eom's)}} 
   - \frac{i}{4\pi}\int_{\Sigma'}^{\Sigma''} d^3x\,A_\nu \delta( \sqrt{h}\, 
   {\bar
    u}_\mu
   F^{\mu\nu})  + \frac{i}{4\pi} \int_{\cal T} d^3x\, A_\nu \delta( 
   \sqrt{-\gamma} 
   \, n_\mu F^{\mu\nu}) \nonumber\\ 
   & = & {\hbox{(eom's)}} 
   - \frac{i}{4\pi}\int_{\Sigma'}^{\Sigma''} d^3x\, A_{\underline i} 
   \delta( \sqrt{h}\,
   {\bar u}_\mu
   F^{\mu\nu} h_\nu^{\underline i})  
   + \frac{i}{4\pi}\int_{\cal T} d^3x\, A_i \delta( \sqrt{-\gamma}\, n_\mu 
   F^{\mu\nu} 
   \gamma_{\nu}^{i}) \ .\eqnum{9}
\end{eqnarray} 
(The derivation uses the results $\delta (\sqrt{-h/g}\, {\bar u}_\mu) = 0$ 
and 
$\delta (\sqrt{\gamma/g}\, n_\mu) = 0$.) 

The electric field on $\Sigma'$ or $\Sigma''$, as measured by the observers
${\bar u}^\mu$, is 
\begin{equation} 
   E^{\underline i} = - {\bar u}_\mu F^{\mu\nu} h_\nu^{\underline i} 
   \ .\eqnum{10} 
\end{equation} 
The result (9) shows that in the variational principle for $S_2$ the 
electric field
$E^{\underline i}$ is fixed on $\Sigma'$ and $\Sigma''$. (Again, it is 
assumed that
the induced metric on the boundary of $\cal M$ is fixed.) Now consider the 
boundary
element $\cal T$. Equation (9) shows that $n_\mu F^{\mu\nu} 
\gamma_{\nu}^{i}$ is
fixed on $\cal T$. Using the decomposition $F^{\mu\nu} = 2
u^{{\scriptscriptstyle [}\mu} E^{\nu{\scriptscriptstyle ]}} +
\epsilon^{\mu\nu\rho\sigma} u_\rho B_\sigma$ for the field strength tensor, 
we find 
\begin{equation} 
   n_\mu F^{\mu\nu} \gamma_{\nu}^{i} = - (E^\vdash) u^i + 
   \epsilon^{ab}\sigma^i_a B_b \eqnum{11} 
\end{equation} 
where $E^\vdash = E^\mu n_\mu$. 
Therefore the fixed  data for $S_2$ includes of the component $E^\vdash$ 
of the electric field orthogonal to the system boundary $\cal B$ and the 
components
$B^a$ of the magnetic field tangent to $\cal B$. The electric charge $Q_e$ 
for the
system, defined by 
\begin{equation} 
   4\pi Q_e = \int_{\cal B} {\mathbf{{}^*F}}= \int_{\cal B}
d^2z\sqrt{\sigma} E^\vdash 
   \ ,\eqnum{12} 
\end{equation}
is fixed as well. In this paper only the
sourceless Maxwell equations are considered; however, a nonvanishing 
electric charge (like magnetic charge) can arise through the presence of 
charged black holes. 

Now turn to the derivation of the duality relationship 
$S_{1}[\tilde g,\tilde F] = S_{2}[\tilde g,-{}^*{\tilde F}]$. Consider a 
solution 
$g_{\mu\nu} = {\tilde g}_{\mu\nu}$, $A_\mu = {\tilde A}_\mu$ of the
classical Einstein--Maxwell equations. The potential yields the field 
strength 
${\tilde F}_{\mu\nu} = 2\partial_{{\scriptscriptstyle [}\mu} {\tilde
A}_{\nu{\scriptscriptstyle ]}}$. The dual classical solution is defined by 
$g_{\mu\nu} = {\tilde g}_{\mu\nu}$, 
$A_\mu = {}^{\scriptscriptstyle D}\!{\tilde
A}_\mu$, where  the potential ${}^{\scriptscriptstyle D}\!{\tilde A}_\mu$ 
yields the
dual of ${\tilde F}_{\mu\nu}$; that is,  ${}^*{\tilde F}_{\mu\nu} =
2\partial_{{\scriptscriptstyle [}\mu} {}^{\scriptscriptstyle D}\!{\tilde
A}_{\nu{\scriptscriptstyle ]}}$.\footnote{In differential forms notation, 
${\tilde
{\mathbf{F}}} = d{\tilde {\mathbf{A}}}$. Since
${\tilde g}_{\mu\nu}$, ${\tilde A}_\mu$ satisfies the Maxwell equations, 
we have
$d{}^*{\tilde {\mathbf{F}}} = 0$. Then the Poincar\'e lemma guarantees the 
local
existence of
${}^{\scriptscriptstyle D}\!{\tilde {\mathbf{A}}}$ such that ${}^*{\tilde
{\mathbf{F}}} =  d{}^{\scriptscriptstyle D}\!{\tilde
{\mathbf{A}}}$.}  We can evaluate the action $S_1[g,F]$ at the solution 
${\tilde
g}_{\mu\nu}$,
${\tilde A}_\mu$ simply by substituting ${\tilde g}$ for $g$ and
${\tilde F}$ for $F$ into Eq.~(1). When the action $S_2[g,F]$ is evaluated 
at the dual
solution 
${\tilde g}_{\mu\nu}$, ${}^{\scriptscriptstyle D}\!{\tilde A}_\mu$, we 
find that the
term in Eq.~(8) involving 
$\partial_\mu (\sqrt{-{\tilde g}}\, {}^*{\tilde F}^{\mu\nu})$ vanishes
identically.\footnote{The equation $\partial_\mu (\sqrt{-{\tilde g}}\, 
{}^*{\tilde
F}^{\mu\nu}) = 0$ is equivalent to $d{\tilde {\mathbf{F}}} = 0$, which 
holds because 
${\tilde {\mathbf{F}}}= d{\tilde {\mathbf{A}}}$.} Then using the result 
${}^*{\tilde
F}^{\mu\nu}\, {}^*{\tilde F}_{\mu\nu} = - {\tilde F}^{\mu\nu} 
{\tilde F}_{\mu\nu}$,
we obtain the relationship 
\begin{equation} 
   S_1[{\tilde g},{\tilde F}] = S_2[{\tilde g},-{}^*{\tilde F}] =
   -\frac{i}{16\pi}\int_{\cal M} d^4X \sqrt{-{\tilde g}} 
    {\tilde F}^{\mu\nu} {\tilde F}_{\mu\nu} \ .\eqnum{13}
\end{equation} 
As discussed in the introduction, this result shows that the rate of 
creation of magnetically charged black holes under a given set of boundary 
conditions is equal to the rate of creation of electrically charged black 
holes under the dual set of boundary conditions. 

It should be noted that the analysis presented here applies to
both  Lorentzian and Euclidean geometries, as well as to more general 
complex
geometries.  In particular, the relationship (13) holds for any metric 
signature. 
Of course, $\sqrt{-{\tilde g}}$ is imaginary if ${\tilde g}_{\mu\nu}$ is
Euclidean.
Likewise,  the  volume form
$\epsilon_{\mu\nu\rho\sigma}$, which appears in the definition of the dual 
field
strength ${}^*{\tilde F}_{\mu\nu}$, is imaginary if ${\tilde g}_{\mu\nu}$ 
is Euclidean. However, in general, the analysis is completely 
self--consistent and there is no need to construct separate arguments for 
the Lorentzian and Euclidean cases. 

As a final remark, let me point out that no action principle for 
electrodynamics has
the full electric field, $E^\vdash$ and $E^a$, for its boundary conditions
on ${\cal T}$. Likewise, no action principle has the full magnetic field, 
$B^\vdash$ and
$B^a$, for its boundary conditions on ${\cal T}$. The easiest way to see 
this is to
recognize that fixation of $E^1\sim \partial_1 A_0 - \partial_0 A_1$, 
$E^2 \sim
\partial_2 A_0 - \partial_0 A_2$, and $E^3\sim \partial_3 A_0 - 
\partial_0 A_3$ on
${\cal T}$ is analogous to fixation of $B^2\sim \partial_1 A_3 - 
\partial_3 A_1$,
$B^1\sim \partial_2 A_3 - \partial_3 A_2$, and $E^3\sim \partial_0 A_3 - 
\partial_3
A_0$ on $\Sigma'$ and $\Sigma''$. Here, for simplicity, I have assumed 
flat Minkowski
spacetime with coordinates $X^\mu$. Also, ${\cal T}$ coincides with an 
$X^3 = {\rm
const}$ surface and $\Sigma'$ and $\Sigma''$ coincide with $X^0 = 
{\rm const}$
surfaces. Now, the quantities fixed at the endpoints in time (that is, at 
$\Sigma'$ and
$\Sigma''$) are the canonical coordinates. Therefore the fixed quantities 
must have
vanishing Poisson brackets among themselves. This is not the case for 
$B^2$, $B^1$, and
$E^3$. From the point of view of canonical transformations, this argument 
shows that
there is no boundary term which, when added to the action $S_1$ (or $S_2$), 
leads to
$B^2$, $B^1$, and $E^3$ as the fixed data on $\Sigma'$ and 
$\Sigma''$.\footnote{From the quantum mechanical point of view, this 
argument shows 
that a specification of $B^2$, $B^1$, and
$E^3$ would violate the uncertainty principle.} Analogously, we find that 
there is no
boundary term which, when added to the action $S_1$ (or $S_2$), leads to 
$E^\vdash$,
$E^a$ as fixed data on
${\cal T}$. A similar argument shows that fixation of $B^\vdash$, $B^a$ is 
not possible on ${\cal T}$. 

%%%%%%%%%%%%%%%%%%%%%%%%%%%%%%%%%%%%%%%%%%%%%%%%%%%%%%%%%%%%%%%%%%%%%%%%%
%%%%%%%%%%%%%%%%%      REFERENCES       %%%%%%%%%%%%%%%%%%%%%%%%%%%%%%%%%
%%%%%%%%%%%%%%%%%%%%%%%%%%%%%%%%%%%%%%%%%%%%%%%%%%%%%%%%%%%%%%%%%%%%%%%%%
 
\end{document}